\documentclass[twocolumn,showpacs,preprintnumbers,amsmath,amssymb,prd]{revtex4}
\usepackage{graphicx,psfrag}
\usepackage{dcolumn}
\usepackage{bm}

\newcommand{\ie}{\textit{i.e.}}
\newcommand{\rmi}{{\rm i}}
\newcommand{\rmd}{{\rm d}}

\begin{document}

\title{BFKL equation at finite temperature}

\author{Kazuaki Ohnishi}
 \email{kohnishi@phya.yonsei.ac.kr}
\author{Su Houng Lee}
 \email{suhoung@phya.yonsei.ac.kr}

\affiliation{Institute of Physics and Applied Physics, Yonsei University,
Seoul 120-749, Korea}
\date{\today}

\begin{abstract}
We consider the Color Glass Condensate (CGC) at finite temperature,
which would be relevant to the initial condition of relativistic heavy
ion collisions and the energy loss of energetic partons in the quark-gluon
plasma. In the weak source approximation, we derive the thermal BFKL equation.
We find that the thermal effect shows up as a Bose enhancement of the soft
gluon emission if the temperature is high enough to match the light-cone
energy of the soft gluons. This suggests that the saturation regime could be
reached sooner than in the vacuum.
\end{abstract}

\pacs{11.10.Wx, 12.38.Cy, 12.38.Mh}
\maketitle

\section{Introduction}
The Relativistic Heavy Ion Collider  (RHIC) at BNL has revealed various
interesting natures of high energy QCD matter which would be the quark-gluon
plasma (QGP). One of the findings is that the expanding matter can be described
as a perfect fluid without dissipations
\cite{Teaney:2000cw,Kolb:2000fh,Teaney:2003kp,Hirano:2005wx}.
Even surprising would be the early thermalization. Hydrodynamic analyses
suggest that the thermalization takes place in less than 1 fm/$c$ after
the collision \cite{Kolb:2000fh}. In the future Large Hadron
Collider (LHC) experiments at CERN, where higher energy collisions are possible,
we expect an even earlier thermalization, that could be almost instantaneous
after the collision moment \cite{Blaizot:1987nc}. It is important issues to
elucidate the mechanism of the early thermalization
\cite{Hwa:1985tv,Shuryak:1992wc,Baier:2000sb,Arnold:2003rq,Mrowczynski:1988dz,
Romatschke:2003ms,Romatschke:2006nk,Gelis:2006ks},
as well as to determine the initial condition to be used as an input of
hydrodynamical simulations
\cite{Eskola:1988hp,Blaizot:1987nc,Kovner:1995ja,Kovchegov:1997ke,
Gyulassy:1997vt, Kovchegov:2000hz,Krasnitz:1998ns}. Another
discovery to be mentioned would be the jet quenching. QGP has an enormous
stopping power for energetic partons, for which full theoretical understanding
has not been achieved.

On the other hand, small $x$ physics in high energy hadron scatterings is
one of the fascinating areas in QCD physics. The gluon distribution in the small
$x$ region, which is responsible for the high energy hadronic scattering
amplitude, is determined by the BFKL equation
\cite{Lipatov:1976zz,Kuraev:1977fs,Balitsky:1978ic}.
It is known, however, that the BFKL equation predicts too many
gluons as we go to smaller $x$, resulting in unitarity violation of the
scattering cross-section. We need some saturation mechanism to attenuate
the growth of the small $x$ gluon number
\cite{Gribov:1984tu,Mueller:1985wy,Blaizot:1987nc,Mueller:1999wm,Iancu:2001md}.
The Color Glass Condensate (CGC) is one of the formalisms which successfully
implements the saturation
\cite{McLerran:1993ni,Ayala:1995kg,Kovchegov:1996ty,Jalilian-Marian:1996xn,
Jalilian-Marian:1997jx,Jalilian-Marian:1997gr,Jalilian-Marian:1997dw,
Kovner:2000pt,Iancu:2000hn,Iancu:2001ad,Ferreiro:2001qy}.
The idea of CGC is simple and clear. We assume the presence of recoilless large
$x$ partons which travel at the speed of light and stay on the
light-cone. These large $x$ partons act as static color sources, which radiate
smaller $x$ partons. And the emitted gluons radiate further smaller $x$ partons.
At some point, however, the population becomes so large that the recombination
between gluons begins to be effective, resulting in the saturation in the small
$x$ region. The state in the saturated regime, where the gluon field can be
treated as classical, is called the Color Glass Condensate. Mathematically,
the successive emission of smaller and smaller $x$ gluons is formulated as
a Renormalization Group (RG) equation of the effective action. Within the
formalism, the BFKL equation can be reproduced in the weak field regime.
In the saturation regime, the evolution equation is replaced by the JIMWLK
equation, which is a non-linear extension of the BFKL equation.

What we intend to do in this paper is to extend the CGC formalism to finite
temperature and to derive the thermal BFKL equation.
There are two reasons for doing that.   (1) As we have mentioned,
the matter reaches thermalization just after the collision, through
some very complicated processes.  Actually, the mechanism of the early
thermalization is not fully understood so far. It may be that the mechanism
should be verified within the CGC formalism itself
\cite{Romatschke:2006nk,Gelis:2006ks}. For our purpose, let us assume
the early thermalization as an empirical fact. On the other hand,
in the Bjorken picture of
the ultra-relativistic nuclear collision \cite{Bjorken:1982qr}, the large $x$
partons including valence partons can go through the other nucleus
without damage, and remain almost on the light-cone.
Then, just after the collision, the large $x$ partons would radiate smaller
$x$ partons in the medium at finite temperature. This gluon distribution,
as is determined by the thermal BFKL or JIMWLK equation, would give us
the initial condition for the hydrodynamical expansion that follows. It may be
that CGC is reconstructed at that moment, which is not necessarily identical
with that being present before the collision \cite{Asakawa:2006tc}. In fact,
in Refs.\ \cite{Hirano:2004rs,Adil:2005bb,Hirano:2005xf}, it was shown
that CGC can become an appropriate initial condition.

(2) The other motivation is related to the jet quenching
\cite{Liu:2006ug,Casalderrey-Solana:2007sw}. As is discussed in
Ref.\ \cite{Casalderrey-Solana:2007sw}, the energetic parton going through
QGP can be regarded as a collision of the parton and the QGP with the large
center-of-mass energy. Just as in the high energy hadronic scattering where
the hadrons exhibit the saturation, the QGP would have saturated gluons as
a consequence of radiation by the thermal quarks and gluons. Those extra
emitted gluons could explain the observed large stopping ability of QGP.
We notice that this argument is in a frame where the thermal bath of QGP
is moving at high speed. In the rest frame of QGP, we have another picture,
which should be equivalent to the physics in the QGP moving frame.  Here the
energetic parton radiates soft gluons inside the medium at finite temperature
that is at rest. Those radiated  soft gluons, the distribution of which could
be determined by the thermal BFKL equation derived in the CGC formalism, would
be responsible for the parton's energy loss in QGP.

Motivated by these reasons, we will extend the CGC formalism to finite temperature
to derive the thermal BFKL equation.

The organization of this paper is as follows. In Sec.\ \ref{finite T},
we present the gluon propagator in the light-front field theory at finite
temperature. In Sec.\ \ref{thermal CGC}, we discuss the CGC formalism
at finite temperature. In Sec.\ \ref{thermal BFKL}, we derive the thermal
BFKL equation. Summary is given in Sec.\ \ref{summary}.

Throughout this paper, we use the same convention of the light-cone variable
and the metric and so on as in Ref.\ \cite{Iancu:2000hn}.

\section{Light-front field theory at finite temperature}
\label{finite T}

Let us begin with basic concepts of (conventional) thermal field theory
\cite{Landsman:1986uw,Kapusta}. All the thermodynamic quantities can be
derived from the partition function $Z={\rm Tr}\exp(-\beta H)$ with the
Hamiltonian $H$ and the inverse temperature $\beta$, which can be rewritten
in the path integral form. There are two formalisms available: The
imaginary-time formalism and real-time formalism. In the imaginary-time
formalism, the (anti-)periodic boundary condition with respect to the
Euclidean time is imposed on the Bose (Fermion) fields. This results in
the discrete Matsubara frequency and the summation over it is involved.
In the real-time formalism, on the other, we deal with the real (Minkowski)
time directly, at the cost of extending the theory to the complex time.
The complex time path has some arbitrariness under some constraint, and
one possible choice, which is usually employed, is the one composed of
the three paths: $C_{+}=[-\infty:\infty]$,
$C_{-}=[\infty-\rmi\eta:-\infty-\rmi\eta]$ and
$C_{\rm vertical}=[-\infty-\rmi\eta:-\infty-\rmi\beta]$, among which the
last piece eventually decouples from the theory. Corresponding to the
two time paths of $C_{+}$ and $C_{-}$, we have two kinds of interaction
vertex, and the propagator becomes a two-by-two matrix form.

CGC is formulated in the light-front field theory
\cite{Burkardt:1995ct,Brodsky:1997de}, so that we need the light-front field
theory at finite temperature when considering the thermal CGC. The light-front
thermal field theory was initiated in Ref.\ \cite{Brodsky:2001bx} and
formulated in Refs.\
\cite{Alves:2002tx,Weldon:2003uz,Das:2003mf,Kvinikhidze:2003wc,Raufeisen:2004dg},
In the light-front thermal field theory, the (imaginary and real) time
variable now refers to the light-cone time $x^{+}$. It is noted that the
partition function is given by
${\rm Tr}[ \exp \{ -( P^{+}+P^{-})/ \sqrt{2} T \}]$ rather than
${\rm Tr}[ \exp \{ -P^{-}/T \}]$ where $P^{-}$ is  the light-front Hamiltonian.
Also, the theory has non-trivial dependence on the frame we choose.
It is argued that it is impossible to formulate the theory in the frame
where the heat bath moves at the speed of light. The theory formulated in
the rest frame of the heat bath turns out to be the simplest. In the
following, we will work in the rest frame of the heat bath, which is also
the physically relevant frame for describing an actual heavy ion collision
that we are interested in.

For our purpose, we employ the real-time formalism rather than that based on
the imaginary-time. We have two reasons:
(1) As is argued in Ref.\ \cite{Iancu:2000hn}, CGC is formulated properly
on the complex time contour $C=C_{+}\cup C_{-}$. In the real-time formalism,
it is possible to obtain the finite temperature theory just by adding
the path $C_{\rm vertical}$ to $C$.
(2) When we derive the RG equation which gives us the BFKL and JIMWLK
equations, we have to perform the shell-momentum integration over the
energy variable $p^-$. In the imaginary-time formalism, we have to do the
analytic continuation from the discrete Matsubara frequency to the
continuous energy variable, while in the real-time formalism, we can treat
the continuous $p^{-}$ directly.

The bare real-time thermal gluon propagator in the light-front theory
$G^{\mu\nu}_{0(\alpha\beta)}$ was given in Ref.\ \cite{Das:2003mf}, where
$\mu, \nu =\pm,i$ ($i=1,2$) are the Lorentz indices and $\alpha,\beta=\pm$
are the indices associated with the contours $C_{+}$ and $C_{-}$, and
the color indices are suppressed. We put the indices $\alpha,\beta$
in parentheses in order to distinguish them from the Lorentz indices
which also take the values of $\pm$. In Appendix \ref{propagator},
we explicitly write down the non-zero components of
$G^{\mu\nu}_{0(\alpha\beta)}$ in the light-cone gauge ($A_{a}^{+}=0$),
which will be used in the following discussions.

As is emphasized in Ref.\ \cite{Iancu:2000hn}, it is important to specify
the prescription for the pole in $1/p^{+}$. If we use an inappropriate
prescription, we are in a trouble as the soft gluon emission becomes
non-local: There appear smaller $x$ gluons in negative $x^-$ even if
the color source lies in positive $x^-$. This has a serious consequence
especially for deriving the JIMWLK equation. We use the retarded
prescription as in Ref.\ \cite{Iancu:2000hn}, in which all the radiated
gluons come in positive $x^-$ definitely as it should be.

It is noted that there is still an arbitrariness for the pole prescription
for the $G_{(+-)}$ and $G_{(-+)}$ components even if $G_{(++)}(=G_{(--)}^{*})$
is specified for which we use the retarded prescription. In this paper,
we take the same prescription for the $G_{(+-)}$ and $G_{(-+)}$ components
as for $G_{(++)}$, as is shown in Appendix \ref{propagator}. In fact,
Ref.\ \cite{Landshoff:1992ne} proposes a more symmetric prescription
in which, for instance, the $G^{i-}_{0(+-)}$ component has the form of
\begin{align*}
\rmi G^{i-}_{0(+-)}(p) =&
\left[
\theta\left(-p^0\right)+ n_{\rm B}\left(\left|p^{0}\right|\right)
\right]
\\
&\times\rmi
\left(
G_{0}(p)\frac{p^i}{p^{+}+\rmi\epsilon}
-G_{0}^{*}(p)\frac{p^i}{p^{+}-\rmi\epsilon}
\right),
\end{align*}
while in our prescription we have;
\begin{align*}
\rmi G^{i-}_{0(+-)}(p) =& \frac{p^i}{p^{+}+\rmi\epsilon}2\pi
\left[
\theta\left(-p^0\right)+ n_{\rm B}\left(\left|p^{0}\right|\right)
\right]
\delta\left(p^{2}\right)
\\
=&
\left[
\theta\left(-p^0\right)+ n_{\rm B}\left(\left|p^{0}\right|\right)
\right]
\\
&\times\rmi \frac{p^i}{p^{+}+\rmi\epsilon}\left(G_{0}(p)-G_{0}^{*}(p)\right).
\end{align*}
In the symmetric prescription, the pole prescription is replaced by its
Hermite conjugate when it is multiplied by $G_{0}^{*}$.
It can be shown explicitly that in the symmetric prescription, there arises
a contribution of the gluon distribution in negative $x^{-}$, while
in our prescription, the radiated gluons appear only in positive $x^{-}$.
Thus our prescription is the appropriate one for calculations of CGC.

\section{Color Glass Condensate at finite temperature}
\label{thermal CGC}

In this section, we discuss how the CGC formalism is extended to finite
temperature. Let us begin with recapitulating the theory at zero
temperature \cite{Iancu:2000hn,McLerran:2001sr}. CGC is the high gluon
density matter that appears in the high energy limit of hadrons and
is relevant to the hadron scatterings in the Regge kinematics.
Specifically, we can imagine the high energy hadron-hadron
(or nucleus-nucleus) collision or Deep Inelastic Scattering (DIS) where
a high energy virtual photon collides with a hadron.
The generating (or partition) functional for CGC is given by
\begin{equation}
\mathcal{Z}[j]=\int\mathcal{D}\rho W_{\Lambda}[\rho]Z_{\Lambda}^{-1}[\rho]
\int^{\Lambda}\mathcal{D}A_{a}^{\mu}\delta\left(A_{a}^{+}\right)
{\rm e}^{\rmi S[A, \rho ]-\rmi\int j\cdot A},
\end{equation}
which contains the generating functional for the soft gluon field $A^{\mu}$
at a fixed color charge distribution $\rho$;
\begin{equation}
Z_{\Lambda}[\rho, j]=
Z_{\Lambda}^{-1}[\rho]
\int^{\Lambda}\mathcal{D}A_{a}^{\mu}\delta\left(A_{a}^{+}\right)
{\rm e}^{\rmi S[A, \rho ]-\rmi\int j\cdot A}
\end{equation}
with $Z_{\Lambda}[\rho]\equiv Z_{\Lambda}[\rho, j=0]$. The frozen color charge
$\rho$ simulates the effect of the fast partons with momenta
$|p^{+}|>\Lambda^{+}$, and the charge distribution is specified by the weight
function $W_{\Lambda}[\rho]$ into which all the dynamics of the fast modes is
encoded. Let us recall how we can calculate the gluon distribution function
within the formalism. There are two ingredients:

\begin{description}
\item[(i)] We solve the classical Yang-Mills equation to some specific charge
distribution, assuming the presence of the strong gluon field at the small $x$
region of interest.

\item[(ii)] The gluon distribution function is obtained by averaging the
classical solution $\mathcal{A}^{i}(\vec{x}=(x^{-},x_{\perp}))[\rho]$
we have obtained over $\rho$ with the weight function $W_{\Lambda}[\rho]$.
\end{description}

We note that the classical solution $\mathcal{A}^{i}(\vec{x})[\rho]$ we need
is a static, time-independent one, which is necessary for computing the gluon
distribution function.

If we lower the cutoff $\Lambda^{+}$ by integrating out the semi-fast gluons
$a^{\mu}$ ($b\Lambda^{+}<|p^{+}|<\Lambda^{+}$), their dynamics is renormalized
into $W_{b\Lambda}[\rho]$ to give the RG equation for $W_{\Lambda}[\rho]$.
Convoluted with the classical solution, this RG equation for
$W_{\Lambda}[\rho]$ finally gives us the evolution equation for the gluon
distribution function, that is, the BFKL and JIMWLK equations. The RG
equation for $W_{\Lambda}[\rho]$ reads
\begin{align}
&\frac{\partial W_{\tau}[\rho]}{\partial\tau}
\nonumber\\
&=\alpha_{\rm s}
\left\{
\frac{1}{2}\frac{\delta^2}{\delta\rho_{\tau}(x)\delta\rho_{\tau}(y)}
\left[W_{\tau}\chi_{xy}\right]
-\frac{\delta}{\delta\rho_{\tau}(x)}
\left[W_{\tau}\sigma_{x}\right]
\right\},
\label{eq:RG}
\end{align}
where
\begin{align}
\hat{\sigma}_{a}(\vec{x})
&\equiv\langle\delta\rho_{a}(x)\rangle,
\\
\hat{\chi}_{ab}(x,y)
&\equiv\langle\delta\rho_{a}(x)\delta\rho_{b}(y)\rangle,
\end{align}
and $\tau=\ln(P^{+}/\Lambda^{+})=\ln(1/x)$ is the rapidity variable.
$\hat{\sigma}(\vec{x})$ and $\hat{\chi}(x,y)$ are the one- and
two-point functions, respectively, of the induced color charge which is
generated by the semi-fast gluon fluctuation and is defined by
\begin{align}
\delta J_{\mu}^{a}(x)
\equiv & -\left.\frac{\delta S}{\delta A_{a}^{\mu}(x)}\right|_{\mathcal{A}+a}
\nonumber\\
\approx &
-\left.\frac{\delta^{2}S}{\delta A_{a}^{\mu}(x)\delta A_{b}^{\nu}(y)}
\right|_{\mathcal{A}}a_{b}^{\nu}(y)
\nonumber\\
&-\frac{1}{2}\left.\frac{\delta^{3}S}{\delta A_{a}^{\mu}(x)
\delta A_{b}^{\nu}(y)\delta A_{c}^{\lambda}(z)}
\right|_{\mathcal{A}}a_{b}^{\nu}(y)a_{c}^{\lambda}(z)
\label{induced charge}
\end{align}
with $\mu=-$.
The derivation of the RG equation is reduced to computation of
$\hat{\sigma}(\vec{x})$ and $\hat{\chi}(x,y)$.

Now we consider the extension to finite temperature of the formalism.
First, we note that the classical solution at finite temperature is
identical with that at zero temperature: The thermal effect does not
alter the classical solution within the classical approximation.
This can be seen as follows. The thermal effect in finite temperature
field theory is brought about by the modification of the time contour
on which the theory lives. As we noticed, the classical solution needed
in the CGC formalism is the time-independent one. Thus, the classical
solution at finite temperature is provided by the same one as
at zero temperature.

All the thermal fluctuations are contained in the correlators
$\hat{\sigma}(\vec{x})$ and $\hat{\chi}(x,y)$, the evaluation of which
is all we have to do. The action in the real time formalism at finite
temperature is given by
\begin{align}
S[A,\rho] &=S_{\rm YM}+S_{W},
\\
S_{\rm YM} &=-\int_{C}\rmd^{4}x\frac{1}{4}F_{\mu\nu}^{a}F_{a}^{\mu\nu},
\\
S_{W}[A^{-},\rho] &= \frac{\rmi}{gN_{\rm c}}
\int \rmd^{3}\vec{x} {\rm Tr}
\left[\rho(\vec{x})W_{C}[A^{-}](\vec{x})\right]
\end{align}
with the time-ordered Wilson line
\begin{equation}
W_{C}(\vec{x})={\rm T}_{C}\exp\left\{\rmi g\int_{C}\rmd z A^{-}(z,\vec{x})
\right\},
\end{equation}
where $C$ denotes the complex time contour given by
$C=C_{+}\cup C_{-}\cup C_{\rm vertical}$, for which the last piece is
eventually decoupled from the theory \cite{Landsman:1986uw}. With this
action, we can calculate the induced charge $\delta\rho$ defined in
Eq.\ (\ref{induced charge}), which reads
\begin{equation}
\delta\rho_{a}(x) = \delta\rho_{a}^{(1)}(x)+\delta\rho_{a}^{(2)}(x),
\end{equation}
where
\begin{widetext}
\begin{align}
\delta\rho_{a}^{(1)}(x) =&
-2\rmi g\mathcal{F}_{ac}^{+i}(\vec{x})a^{ic}(x)
\nonumber\\
&-g\rho_{ac}(\vec{x})
\left[
\int_{-\infty}^{\infty}\rmd y_{(+)}^{+}
\langle x^{+}|{\rm PV}\frac{1}{\rmi \partial_{y}^{-}}|y^{+}\rangle
a^{c-}(y_{(+)}^{+},\vec{x})
+\frac{\rmi}{2}\int_{-\infty}^{\infty}\rmd y_{(-)}^{+}
a^{c-}(y_{(-)}^{+},\vec{x})
\right],
\end{align}
\begin{align}
\delta\rho_{a}^{(2)}(x) =&
g^{2}f^{abc}\left[\partial^{+}a_{i}^{b}(x)\right]a_{i}^{c}(x)
\nonumber\\
&-\frac{g^2}{N_{\rm c}}\rho^{b}(\vec{x})
\left[
\int_{-\infty}^{\infty}\rmd y_{(+)}^{+}
\int_{-\infty}^{\infty}\rmd z_{(+)}^{+}
a^{c-}(y_{(+)}^{+},\vec{x})a^{d-}(z_{(+)}^{+},\vec{x})
\right.
\nonumber\\
&\hspace{20mm}\times\left\{
\theta(x^{+}-y_{(+)}^{+})\theta(y_{(+)}^{+}-z_{(+)}^{+})
{\rm Tr}(T^{a}T^{c}T^{d}T^{b})
+\theta(y_{(+)}^{+}-z_{(+)}^{+})\theta(z_{(+)}^{+}-x^{+})
{\rm Tr}(T^{a}T^{b}T^{c}T^{b})
\right.
\nonumber\\
&\hspace{25mm}\left.
+\theta(y_{(+)}^{+}-x^{+})\theta(x^{+}-z_{(+)}^{+})
{\rm Tr}(T^{a}T^{d}T^{b}T^{c})
\right\}
\nonumber\\
&\hspace{20mm}-\int_{-\infty}^{\infty}\rmd y_{(-)}^{+}
\int_{-\infty}^{\infty}\rmd z_{(+)}^{+}
a^{c-}(y_{(-)}^{+},\vec{x})a^{d-}(z_{(+)}^{+},\vec{x})
\theta(z_{(+)}^{+}-x^{+}){\rm Tr}(T^{a}T^{b}T^{c}T^{d})
\nonumber\\
&\hspace{20mm}+
\int_{-\infty}^{\infty}\rmd y_{(-)}^{+}
\int_{-\infty}^{\infty}\rmd z_{(-)}^{+}
a^{c-}(y_{(-)}^{+},\vec{x})a^{d-}(z_{(-)}^{+},\vec{x})
\theta(z_{(-)}^{+}-y_{(-)}^{+}){\rm Tr}(T^{a}T^{b}T^{c}T^{d})
\nonumber\\
&\hspace{20mm}\left.
-\int_{-\infty}^{\infty}\rmd y_{(-)}^{+}
\int_{-\infty}^{\infty}\rmd z_{(+)}^{+}
a^{c-}(y_{(-)}^{+},\vec{x})a^{d-}(z_{(+)}^{+},\vec{x})
\theta(x^{+}-z_{(+)}^{+}){\rm Tr}(T^{a}T^{d}T^{b}T^{c})
\right].
\end{align}
\end{widetext}
It is noted that the terms that involve the integration over the time
variable on $C_{-}$ (\ie, $y_{(-)}^{+}$ and $z_{(-)}^{+}$) represent
the thermal effects, while the other terms are already present at zero
temperature.

Having obtained the charge fluctuation, we can now express
$\hat{\sigma}(\vec{x})$ and $\hat{\chi}(x,y)$ in terms of the full
Feynman propagator $G_{(\alpha\beta)}^{\mu\nu}(x,y)$ of the semi-fast gluons.
To the lowest order of $\alpha_{\rm s}$, we have
\begin{equation}
\hat{\sigma}^{a}(\vec{x})
=\hat{\sigma}_{1}^{a}(\vec{x})+\hat{\sigma}_{2}^{a}(\vec{x}),
\end{equation}
where
\begin{align}
\hat{\sigma}_{1}^{a}(\vec{x})
&=\left.g^{2}\rmi f^{abc}\partial_{y}^{+}
G_{(++)cb}^{ii}(x^{+},\vec{x};y^{+},\vec{y})\right|_{x=y}
\nonumber\\
&=-g^{2}\partial_{y}^{+}
{\rm Tr}\left[
T^{a}G_{(++)}^{ii}(x^{+},\vec{x};y^{+},\vec{y})\right]_{x=y},
\label{eq:sigma_1}
\end{align}
\begin{widetext}
\begin{align}
\hat{\sigma}_{2}^{a}(\vec{x})
=-\rmi\frac{g^2}{N_{\rm c}}\rho^{b}(\vec{x})
&\left[
\int_{-\infty}^{\infty}\rmd y_{(+)}^{+}
\int_{-\infty}^{\infty}\rmd z_{(+)}^{+}
G_{(++)dc}^{--}(z_{(+)}^{+},\vec{x};y_{(+)}^{+},\vec{x})
\right.
\nonumber\\
&\times\left\{
\theta(x^{+}-y_{(+)}^{+})\theta(y_{(+)}^{+}-z_{(+)}^{+})
{\rm Tr}(T^{a}T^{c}T^{d}T^{b})
+\theta(y_{(+)}^{+}-z_{(+)}^{+})\theta(z_{(+)}^{+}-x^{+})
{\rm Tr}(T^{a}T^{b}T^{c}T^{b})
\right.
\nonumber\\
&\hspace{5mm}\left.
+\theta(y_{(+)}^{+}-x^{+})\theta(x^{+}-z_{(+)}^{+})
{\rm Tr}(T^{a}T^{d}T^{b}T^{c})
\right\}
\nonumber\\
&-\int_{-\infty}^{\infty}\rmd y_{(-)}^{+}
\int_{-\infty}^{\infty}\rmd z_{(+)}^{+}
G_{(+-)dc}^{--}(z_{(+)}^{+},\vec{x};y_{(-)}^{+},\vec{x})
\theta(z_{(+)}^{+}-x^{+}){\rm Tr}(T^{a}T^{b}T^{c}T^{d})
\nonumber\\
&+
\int_{-\infty}^{\infty}\rmd y_{(-)}^{+}
\int_{-\infty}^{\infty}\rmd z_{(-)}^{+}
G_{(--)dc}^{--}(z_{(-)}^{+},\vec{x};y_{(-)}^{+},\vec{x})
\theta(z_{(-)}^{+}-y_{(-)}^{+}){\rm Tr}(T^{a}T^{b}T^{c}T^{d})
\nonumber\\
&\left.
-\int_{-\infty}^{\infty}\rmd y_{(-)}^{+}
\int_{-\infty}^{\infty}\rmd z_{(+)}^{+}
G_{(+-)dc}^{--}(z_{(+)}^{+},\vec{x};y_{(-)}^{+},\vec{x})
\theta(x^{+}-z_{(+)}^{+}){\rm Tr}(T^{a}T^{d}T^{b}T^{c})
\right],
\end{align}
\end{widetext}
and
\begin{equation}
\hat{\chi}(\vec{x},\vec{y})
=\hat{\chi}_{1}(\vec{x},\vec{y})+\hat{\chi}_{2}(\vec{x},\vec{y}),
\end{equation}
where
\begin{equation}
\hat{\chi}_{1}(\vec{x},\vec{y})
=4\rmi g^{2}\mathcal{F}^{+i}(\vec{x})G_{(++)}^{ij}(x,y)
\mathcal{F}^{+j}(\vec{y}),
\label{eq:chi_1_full}
\end{equation}
\begin{widetext}
\begin{align}
\frac{1}{g^2}\hat{\chi}_{2}(\vec{x},\vec{y})
=&
\mathcal{F}^{+i}(\vec{x})
\left\{
-2\int\rmd w_{(+)}^{+} G_{(++)}^{i-}(x^{+},\vec{x};w_{(+)}^{+},\vec{y})
\langle w^{+}|{\rm PV}\frac{1}{\rmi \partial_{y}^{-}}|y^{+}\rangle
+\rmi \int\rmd w_{(-)}^{+} G_{(+-)}^{i-}(x^{+},\vec{x};w_{(-)}^{+},\vec{y})
\right\}
\rho(\vec{y})
\nonumber\\
&+\rho(\vec{x})
\left\{
2\int\rmd z_{(+)}^{+}
\langle x^{+}|{\rm PV}\frac{1}{\rmi \partial_{z}^{-}}|z^{+}\rangle
G_{(++)}^{-j}(z_{(+)}^{+},\vec{x};y^{+},\vec{y})
+\rmi\int\rmd z_{(-)}^{+} G_{(-+)}^{-j}(z_{(-)}^{+},\vec{x};y^{+},\vec{y})
\right\}
\mathcal{F}^{+j}(\vec{y})
\nonumber\\
&+\rho(\vec{x})
\left\{
\rmi\int\rmd z_{(+)}^{+}\int\rmd w_{(+)}^{+}
\langle x^{+}|{\rm PV}\frac{1}{\rmi \partial_{z}^{-}}|z^{+}\rangle
G_{(++)}^{--}(z_{(+)}^{+},\vec{x};w_{(+)}^{+},\vec{y})
\langle w^{+}|{\rm PV}\frac{1}{\rmi \partial_{y}^{-}}|y^{+}\rangle
\right.
\nonumber\\
&\hspace{12mm}+\frac{1}{2}
\int\rmd z_{(+)}^{+}\int\rmd w_{(-)}^{+}
\langle x^{+}|{\rm PV}\frac{1}{\rmi \partial_{z}^{-}}|z^{+}\rangle
G_{(+-)}^{--}(z_{(+)}^{+},\vec{x};w_{(-)}^{+},\vec{y})
\nonumber\\
&\hspace{12mm}-\frac{1}{2}
\int\rmd z_{(-)}^{+}\int\rmd w_{(+)}^{+}
G_{(-+)}^{--}(z_{(-)}^{+},\vec{x};w_{(+)}^{+},\vec{y})
\langle w^{+}|{\rm PV}\frac{1}{\rmi \partial_{y}^{-}}|y^{+}\rangle
\nonumber\\
&\hspace{12mm}\left.\left.
+\frac{1}{4}\rmi
\int\rmd z_{(-)}^{+}\int\rmd w_{(-)}^{+}
G_{(--)}^{--}(z_{(-)}^{+},\vec{x};w_{(-)}^{+},\vec{y})
\right\}
\rho(\vec{y})
\right|_{y^{+}=x^{+}+\epsilon}.
\end{align}
\end{widetext}

The diagrammatic representations are the same as given
in Ref.\ \cite{Iancu:2000hn}, although the internal time variables
can now take values on $C_{-}$, giving rise to the additional terms
corresponding to the thermal effects.

\section{BFKL equation at finite temperature}
\label{thermal BFKL}

It has been shown that the CGC formalism at zero temperature reproduces
the BFKL equation in the weak source approximation
\cite{Jalilian-Marian:1997jx}. In the present section, we derive the
thermal BFKL equation within the finite temperature CGC discussed in the
previous section, using the same approximation.  All we have to do is
to evaluate $\hat{\sigma}$ and $\hat{\chi}$ in the approximation,
for which we need the free propagator
$G_{0(\alpha\beta)}^{\mu\nu}$ presented in Appendix \ref{propagator}.
The discussion in this section is proceeded in
parallel with that given in Sec.\ 5 of Ref.\ \cite{Iancu:2000hn}.

\subsection{Evaluation of $\hat{\sigma}$ in the weak source limit}
We first consider $\hat{\sigma}_1$. Because of the color trace in
Eq.\ (\ref{eq:sigma_1}), the vacuum contribution disappears. The
contribution linear in $\rho$ that has logarithmic enhancement is given
by a direct insertion of $\rho$, which reads
\begin{align}
\hat{\sigma}_{1}(\vec{x})
=& -g^{2}\int_{C}\rmd^{4}z\int_{C}\rmd^{4}u
\nonumber\\
&\left.\times
G_{0}^{i-}(x-z)\Pi^{C}(z,u)\partial_{y}^{+}G_{0}^{-i}(u-y)\right|_{y=x},
\label{eq:sigma_1_config}
\end{align}
where
\begin{align}
\Pi_{ab}^{C}(z,u)
&\equiv
\left.
\frac{\delta^{2}S_{W}}{\delta A_{a}^{-}(z)\delta A_{b}^{-}(u)}
\right|_{\mathcal{A}}
\nonumber\\
&=\frac{\rmi}{2}\rho_{ab}\delta(\vec{z}-\vec{u})
\left\{
\theta_{C}(z^{+}-u^{+})-\theta_{C}(u^{+}-z^{+})
\right\}
\end{align}
after the rescaling $\delta\rho\rightarrow\frac{1}{g}\delta\rho$. We note that
because the time contour $C$ has the two pieces, there arise four terms
for $\hat{\sigma}_{1}$ in Eq.\ (\ref{eq:sigma_1_config}).

In the following, we directly calculate the two-dimensional charge density
$\sigma^{a}(x_{\perp})\equiv\int\rmd x^{-}\hat{\sigma}^{a}(\vec{x})$,
which is actually what we need in the RG equation. In momentum space,
Eq.\ (\ref{eq:sigma_1_config}) is written as
(with $\vec{q}=\vec{p}-\vec{k}$)
\begin{widetext}
\begin{align}
\hat{\sigma}_{1}(\vec{x})
=&-\frac{\rmi}{2}g^{2}
\int\frac{\rmd p^{-}}{2\pi}
\int\frac{\rmd^{3} \vec{p}}{(2\pi)^{3}}
\int\frac{\rmd^{3} \vec{k}}{(2\pi)^{3}}
{\rm e}^{-\rmi\vec{q}\cdot\vec{x}}\rho(\vec{q})
\left\{
-2G_{0(++)}^{i-}(p^{-},\vec{p})\left[{\rm PV}\frac{1}{p^-}\right]k^{+}
G_{0(++)}^{-i}(p^{-},\vec{k})
\right.
\nonumber\\
&+2\pi\rmi\delta({p^-})G_{0(++)}^{i-}(p^{-},\vec{p})k^{+}
G_{0(-+)}^{-i}(p^{-},\vec{k})
-2\pi\rmi\delta({p^-})G_{0(+-)}^{i-}(p^{-},\vec{p})k^{+}
G_{0(++)}^{-i}(p^{-},\vec{k})
\nonumber\\
&\left.
+2G_{0(+-)}^{i-}(p^{-},\vec{p})\left[{\rm PV}\frac{1}{p^-}\right]k^{+}
G_{0(-+)}^{-i}(p^{-},\vec{k})
\right\}.
\label{eq:sigma_1_momentum}
\end{align}
\end{widetext}
As in the case of zero temperature, we can find that $k^{+}$ could be
replaced by $p^{+}$ or $(k^{+}+p^{+})/2$, by noting the symmetry of the
integrand under the simultaneous exchange $\vec{k}\leftrightarrow -\vec{p}$
and $p^{-}\leftrightarrow -p^{-}$. If we take the strip restriction on
$p^{-}$ ($\Lambda^{-}<|p^{-}|<\Lambda^{-}/b$), the second and third terms in
the curly brackets in Eq.\ (\ref{eq:sigma_1_momentum}) disappear due to
the delta-function of $p^{-}$. After a straightforward calculation
similar to that at zero temperature, we obtain
\begin{align}
&\sigma_{1}^{a}(x_{\perp})
=-\frac{g^{2}N_{\rm c}}{2\pi}
\int_{\Lambda^{-}}^{\Lambda^{-}/b}\rmd p^{-}\frac{1}{p^{-}}
\left(
1+2n_{\rm B}\left(\frac{p^-}{\sqrt{2}}\right)
\right)
\nonumber\\
&\times\int\frac{{\rmd}^{2}p_{\perp}}{(2\pi)^2}
\int\frac{{\rmd}^{2}k_{\perp}}{(2\pi)^2}
{\rm e}^{\rmi q_{\perp}\cdot x_{\perp}}
\frac{p_{\perp}\cdot k_{\perp}}{p_{\perp}^{2}k_{\perp}^{2}}
\rho^{a}(q_{\perp}),
\label{eq:sigma_1_final_config}
\end{align}
where $\rho^{a}(q_{\perp})\equiv\rho^{a}(q^{+}=0,q_{\perp})$ and
$n_{\rm B}(x)=1/(\exp({\beta x})-1)$ is the Bose distribution.
It should be mentioned that the strip restriction on $p^{+}$
($b\Lambda^{+}<|p^{+}|<\Lambda^{+}$) leads us to the same result as for
the $p^{-}$ restriction, meaning the implicit symmetry under
$p^{+}\leftrightarrow p^{-}$ of the integrand in
Eq.\ (\ref{eq:sigma_1_momentum}). Fourier transforming
Eq.\ (\ref{eq:sigma_1_final_config}), we finally have
\begin{align}
&\sigma_{1}^{a}(q_{\perp})
=2N_{\rm c}\alpha_{\rm s}
\int_{\Lambda^{-}}^{\Lambda^{-}/b}\rmd p^{-}\frac{1}{p^{-}}
\left(
1+2n_{\rm B}\left(\frac{p^-}{\sqrt{2}}\right)
\right)
\nonumber\\
&\times\rho^{a}(q_{\perp})
\int\frac{{\rmd}^{2}p_{\perp}}{(2\pi)^2}
\left(
\frac{q_{\perp}^2}{2p_{\perp}^{2}(p_{\perp}-q_{\perp})^{2}}
-\frac{1}{p_{\perp}^2}
\right).
\end{align}
We note that at zero temperature, the integral over $p^{-}$ gives
rise to the logarithmic enhancement factor $\ln(1/b)$, while in the
present case at finite temperature, the integral cannot be performed
analytically.

One comment is in order. We could calculate $\hat{\sigma}(\vec{x})$
in the $x^-$ representation, explicitly showing the longitudinal structure.
As we have noted, it is possible to verify that $\hat{\sigma}(\vec{x})$ has
contributions only at positive $x^-$ in the non-symmetric prescription
presented in Appendix \ref{propagator}, while if we use the symmetric
prescription mentioned in Sec.\ \ref{finite T}, it turns out that there
appear contributions at both positive and negative $x^-$. Thus our
non-symmetric prescription is valid for calculations of CGC.

Similarly, we can compute $\sigma_{2}^{a}(q_{\perp})$ to be
\begin{align}
\sigma_{2}^{a}(q_{\perp})
=&2N_{\rm c}\alpha_{\rm s}
\int_{\Lambda^{-}}^{\Lambda^{-}/b}\rmd p^{-}\frac{1}{p^{-}}
\left(
1+2n_{\rm B}\left(\frac{p^-}{\sqrt{2}}\right)
\right)
\nonumber\\
&\times\rho^{a}(q_{\perp})
\int\frac{{\rmd}^{2}p_{\perp}}{(2\pi)^2}
\frac{1}{p_{\perp}^2},
\end{align}
which cancels the tadpole term $1/p_{\perp}^{2}$ in $\sigma_{1}$. Combining
the two pieces, we finally obtain
\begin{align}
\sigma_{a}^{(0)}(q_{\perp})\equiv&\sigma_{1}^{a}+\sigma_{2}^{a}
\nonumber\\
=& N_{\rm c}\alpha_{\rm s}
\int_{\Lambda^{-}}^{\Lambda^{-}/b}\rmd p^{-}\frac{1}{p^{-}}
\left(
1+2n_{\rm B}\left(\frac{p^-}{\sqrt{2}}\right)
\right)
\nonumber\\
&\times\rho^{a}(q_{\perp})
\int\frac{{\rmd}^{2}p_{\perp}}{(2\pi)^2}
\frac{q_{\perp}^2}{p_{\perp}^{2}(p_{\perp}-q_{\perp})^{2}}.
\label{eq:sigma_(0)}
\end{align}

\subsection{Evaluation of $\hat{\chi}$ in the weak source limit}
We start with $\hat{\chi_1}$ given in Eq.\ (\ref{eq:chi_1_full}), which,
in lowest order in $\rho$, reads
\begin{equation}
\left.
\hat{\chi}_{1}^{ab}(\vec{x},\vec{y})
=4\rmi g^{2}\mathcal{F}_{ac}^{+i}(\vec{x})G_{0(++)cd}^{ij}(x,y)
\mathcal{F}_{db}^{+j}(\vec{y})\right|_{y^{+}=x^{+}+\epsilon}.
\end{equation}
As in the zero temperature case, we can manipulate
$G_{0(++)}^{ij}(x,y)$ to be
\begin{align}
&G_{0(++)}^{ij}(x,y)
=-\rmi\delta^{ij}
\int_{b\Lambda^{+}}^{\Lambda^{+}}\frac{\rmd p^{+}}{2\pi}
\frac{1}{2p^{+}}
\nonumber\\
&\times
\int\frac{{\rmd}^{2}p_{\perp}}{(2\pi)^2}
{\rm e}^{\rmi p_{\perp}\cdot(x_{\perp}-y_{\perp})}
\left(
1+2n_{\rm B}
\left(\frac{p^{+}+\frac{p_{\perp}^{2}}{2p^{+}}}{\sqrt{2}}\right)
\right),
\end{align}
where we have imposed the strip restriction on the $p^{+}$ integration,
though it is possible to show that the same result can be obtained
when we take the restriction on $p^{-}$. Up to the leading logarithmic
accuracy (LLA), we can replace $p_{\perp}^{2}$ in the Bose distribution
by $Q_{\perp}^{2}$ which is some typical transverse momentum. Then
the factor in the parentheses can be factorized out of the $p_{\perp}$
integral.
Furthermore, as we will see later, we assume the temperature is comparable
with $P^{-}/x=p^{-}$ with $P^{-}$ being the nucleon light-cone energy,
which is much larger than $p^{+}$ so that the term $p^{+}$ in the argument
of the Bose distribution can be neglected. It is noted that the $p^{+}$ term
is also much less than the term $Q_{\perp}^{2}/2p^{+}\approx p^{-}$
in the argument for the on-shell excitation.
After a simple change of variable, we obtain
\begin{align}
G_{0(++)}^{ij}(x,y)
=&-\rmi\delta^{ij}
\int_{\Lambda^{-}}^{\Lambda^{-}/b}
\frac{\rmd p^-}{2\pi}\frac{1}{2p^-}
\left(
1+2n_{\rm B}\left(\frac{p^-}{\sqrt{2}}\right)
\right)
\nonumber\\
&\times\delta^{(2)}(x_{\perp}-y_{\perp}),
\end{align}
which gives us
\begin{align}
\hat{\chi}_{1}^{ab}(\vec{x},\vec{y})
=&2g^{2}
\int_{\Lambda^{-}}^{\Lambda^{-}/b}
\frac{\rmd p^-}{2\pi}\frac{1}{p^-}
\left(
1+2n_{\rm B}\left(\frac{p^-}{\sqrt{2}}\right)
\right)
\nonumber\\
&\times
\delta^{(2)}(x_{\perp}-y_{\perp})
\mathcal{F}_{ac}^{+i}(\vec{x})\mathcal{F}_{cb}^{+i}(\vec{y}).
\end{align}

In a similar way, $\hat{\chi}_{2}^{ab}$ can be calculated, providing us with
\begin{align}
&\hat{\chi}^{(0)}(\vec{x},\vec{y})
=4\alpha_{\rm s}
\int_{\Lambda^{-}}^{\Lambda^{-}/b}
\rmd p^{-}\frac{1}{p^-}
\left(
1+2n_{\rm B}\left(\frac{p^-}{\sqrt{2}}\right)
\right)
\nonumber\\
&\times
\int\frac{{\rmd}^{2}p_{\perp}}{(2\pi)^2}
\frac{{\rm e}^{\rmi p_{\perp}\cdot(x_{\perp}-y_{\perp})}}{p_{\perp}^{2}}
\left\{
\rho(\vec{x})\rho(\vec{y})
+\rmi\rho(\vec{x})\left(\mathcal{F}^{+i}(\vec{y})p^{i}\right)
\right.
\nonumber\\
&
\left.
-\rmi\left(\mathcal{F}^{+i}(\vec{x})p^{i}\right)\rho(\vec{y})
+p_{\perp}^{2}\mathcal{F}^{+i}(\vec{x})\mathcal{F}^{+i}(\vec{y})
\right\}.
\label{eq:chi_(0)}
\end{align}

\subsection{Thermal BFKL equation}
Now we derive the thermal BFKL equation, using the RG equation (\ref{eq:RG}).
As we have mentioned, contrary to zero temperature, the enhancement factor
$\ln(1/b)$ is not manifest at finite temperature, for which we have to be
careful. If we make the factor explicit, Eq.\ (\ref{eq:RG}) is expressed as
\begin{align}
\frac{\partial W_{\tau}[\rho]}{\partial\tau}
&=\lim_{b\rightarrow 1}\frac{1}{\ln\frac{1}{b}}
\nonumber\\
&\times\left\{
\frac{1}{2}\frac{\delta^2}{\delta\rho_{\tau}(x)\delta\rho_{\tau}(y)}
\left[W_{\tau}\hat{\chi}_{xy}\right]
-\frac{\delta}{\delta\rho_{\tau}(x)}
\left[W_{\tau}\hat{\sigma}_{x}\right]
\right\}.
\label{eq:RG2}
\end{align}
We are interested in the evolution of the diagonal element of the two-point
function $\langle\rho\rho\rangle_{\tau}$,
\begin{equation}
\langle\rho_{a}(k_{\perp})\rho_{a}(-k_{\perp})\rangle_{\tau}
\approx k_{\perp}^{2}
\langle\left|
\mathcal{F}_{a}^{+i}(k_{\perp})
\right|\rangle_{\tau}
\equiv\varphi(x,k_{\perp}^{2}),
\end{equation}
which is called the unintegrated gluon distribution. By multiplying
Eq.\ (\ref{eq:RG2}) by $\int{\rmd}^{3}\vec{x}{\rmd}^{3}\vec{y}
\exp\{-\rmi k_{\perp}\cdot(x_{\perp}-y_{\perp})\}
\rho(\vec{x})\rho(\vec{y})$, and by using
Eqs.\ (\ref{eq:sigma_(0)}) and (\ref{eq:chi_(0)}) for $\sigma^{(0)}$
and $\hat{\chi}^{(0)}$, we obtain
\begin{align}
\frac{\partial}{\partial\tau}\varphi(x,k_{\perp}^{2})
&=\lim_{b\rightarrow 1}\frac{1}{\ln\frac{1}{b}}
\nonumber\\
&\times
\frac{\alpha_{\rm s}N_{\rm c}}{\pi^2}
\int_{\Lambda^{-}}^{\Lambda^{-}/b}
\rmd p^{-}\frac{1}{p^-}
\left(
1+2n_{\rm B}\left(\frac{p^-}{\sqrt{2}}\right)
\right)
\nonumber\\
&\times
\left\{
\int{\rmd}^{2}p_{\perp}
\frac{k_{\perp}^{2}}{p_{\perp}^{2}(p_{\perp}-k_{\perp})^{2}}
\varphi(x,p_{\perp}^{2})
\right.
\nonumber\\
&
\left.
-\frac{1}{2}
\int{\rmd}^{2}p_{\perp}
\frac{k_{\perp}^{2}}{p_{\perp}^{2}(p_{\perp}-k_{\perp})^{2}}
\varphi(x,k_{\perp}^{2})
\right\}.
\label{eq:RG3}
\end{align}

We now discuss how the logarithmic enhancement factor can appear
from the prefactor
\begin{equation*}
\int_{\Lambda^{-}}^{\Lambda^{-}/b}
\rmd p^{-}\frac{1}{p^-}
\left(
1+\frac{2}{{\rm e}^{\frac{p^-}{\sqrt{2}T}}-1}
\right)
\end{equation*}
in Eq.\ (\ref{eq:RG3}).
We note that in the present framework, the temperature $T$ is entirely
a free parameter which can take any value. It is useful to divide three
temperature regimes as follows:
(i) $T\ll\Lambda^{-}/b$,
(ii) $T\sim\Lambda^{-}/b$ and
(iii) $T\gg\Lambda^{-}/b$.
It is noted that as is indicated in Eq.\ (\ref{eq:RG3}), we take the limit
of $b\rightarrow 1$ in the end so that the three regimes will be translated
to
(i) $T\ll\Lambda^{-}$,
(ii) $T\sim\Lambda^{-}$ and
(iii) $T\gg\Lambda^{-}$
after all the calculation.
In the regime (i), the thermal distribution function is exponentially
small in the integral shell so that the thermal effect plays no role up to
LLA and we find that the BFKL equation reduces to that at zero temperature.
In the regime (ii), the distribution function is substantial at the
upper boundary. The thermal effect contributes to the logarithmic
enhancement. This is so because the logarithmic factor comes from the upper
boundary of the integral.
In the regime (iii), the temperature is much higher than the shell energy.
In this case, we find that there appears a power enhancement rather than
the logarithmic one. This means that the BFKL equation which resums the
leading logarithmic terms makes no sense. Physically, the eikonal
approximation breaks down in this regime.

In the following, we restrict ourselves to the regime (ii). Although the
evolution in the regime (iii) is interesting, it is beyond the scope of
this work.

Let us now extract the logarithmic factor. For that purpose, we need to
be careful. As is seen in Eq.\ (\ref{eq:RG3}), we eventually take the limit
of $b\rightarrow 1$ that makes the integral shell infinitesimally small,
leading us to the BFKL equation being a differential equation. Thus, we
are essentially considering the temperature close to $\Lambda^{-}/b$
with $b=1$, that is, $T\approx\Lambda^{-}$. However, in order to extract
the logarithmic factor, we have to temporarily take the limit of
$b\rightarrow 0$ before taking $b\rightarrow 1$, as in the case of the
vacuum BFKL equation \cite{Iancu:2000hn}. This means that the upper
boundary of the integral virtually moves toward infinity. Thus, in this
intermediate stage, it is necessary to impose the constraint of
$T\approx\Lambda^{-}/b$ in order to keep the temperature staying at the
upper boundary, or in order to keep ourselves staying in the regime (ii).
With this constraint, we expand the prefactor as
\begin{align}
\int_{\Lambda^{-}}^{\Lambda^{-}/b}
\rmd p^{-}\frac{1}{p^-}
&\left(
1+\frac{2}{{\rm e}^{\frac{p^-}{\sqrt{2}T}}-1}
\right)
\nonumber\\
&=
a_{-1}\ln\frac{1}{b}+a_{0}+a_{1}b+a_{2}b^{2}+\cdots
\end{align}
with $b\ll 1$. In order to find the coefficient $a_{-1}$, we differentiate
the both side with respect to $b$ to have
\begin{equation*}
1+\frac{2}{{\rm e}^{\frac{\Lambda^-}{\sqrt{2}bT}}-1}
=a_{-1}-a_{1}b-2a_{2}b^{2}+\cdots.
\end{equation*}
If we take $b\rightarrow 0$ with keeping $\Lambda^{-}/bT\approx 1$,
we obtain
\begin{equation}
a_{-1}=1+\frac{2}{{\rm e}^{\frac{\Lambda^-}{\sqrt{2}bT}}-1}.
\end{equation}
Thus, the thermal effect can contribute to the logarithmic enhancement
in the regime (ii).

Now that we have extracted the logarithmic enhancement factor, we turn to
take the limit of $b\rightarrow 1$. We note that in this final stage,
we should abandon the constraint of $T\approx\Lambda^{-}/b$, the purpose
of which is just to extract the logarithmic factor. It is easy to take
the operation of $\lim_{b\rightarrow 1}1/\ln(1/b)$, finally leading us
to the thermal BFKL equation
\begin{align}
\frac{\partial}{\partial\tau}\varphi(x,k_{\perp}^{2})
&=
\frac{\alpha_{\rm s}N_{\rm c}}{\pi^2}
\left(
1+\frac{2}{{\rm e}^{\frac{P^-}{\sqrt{2}Tx}}-1}
\right)
\nonumber\\
&\times
\left\{
\int{\rmd}^{2}p_{\perp}
\frac{k_{\perp}^{2}}{p_{\perp}^{2}(p_{\perp}-k_{\perp})^{2}}
\varphi(x,p_{\perp}^{2})
\right.
\nonumber\\
&
\left.
-\frac{1}{2}
\int{\rmd}^{2}p_{\perp}
\frac{k_{\perp}^{2}}{p_{\perp}^{2}(p_{\perp}-k_{\perp})^{2}}
\varphi(x,k_{\perp}^{2})
\right\},
\end{align}
where we have used the relation $\Lambda^{-}=P^{-}/x$ with $P^{-}$ being
the nucleon light-cone energy. The equation is quite simple in the sense
that the only difference from the BFKL equation at zero temperature is
the Bose enhancement factor in the prefactor. This can be given a natural
physical interpretation that the emission of the softer gluons from
the harder ones is promoted because of the presence of the thermal gluons
all around. We conclude that at finite temperature, the growth of small $x$
gluons can be more rapid and the saturation regime can be reached sooner
than at zero temperature, if the temperature is so high as to be in the
regime (ii), namely $T\sim P^{-}/x$.

Finally, we comment on a different thermal BFKL equation that has been
derived in Ref.\ \cite{deVega:2003ji}.  In their equation, the thermal
modification appears in the kernel in transverse coordinates, while our
equation involves the overall Bose prefactor.  The difference is due to the
fact that they assume the thermalization for modes in the $t$-channel at
the center-mass frame, while  in our framework it is for modes in
the $s$-channel. Their thermal BFKL equation would be useful for investigating
Pomerons at finite temperature as argued in Ref.\ \cite{deVega:2003ji},
and ours would be relevant to the initial state of heavy ion collisions
and the jet quenching.

\section{Summary}
\label{summary}

We have derived the thermal BFKL equation within the formalism of
Color Glass Condensate (CGC). CGC at finite temperature would be relevant
to the initial state of ultra-relativistic heavy ion collisions and to understand
the anomalous cross-section of energetic partons passing through the
quark-gluon plasma (QGP). We have discussed the extension of CGC to
finite temperature and found that the classical solution is not modified
at finite temperature and that all thermal effects are included in the
induced charge correlators $\hat{\sigma}(\vec{x})$ and
$\hat{\chi}(x,y)$ appearing in the renormalization group
equation for the weight function $W_{\Lambda}[\rho]$. In the weak source
approximation, we have derived the thermal BFKL equation by explicitly
evaluating the charge correlators in the real-time formalism. We have
specified the pole prescription for $1/p^{+}$ in the free gluon propagator,
as is presented in Appendix \ref{propagator}. We have employed the
retarded prescription for the $G_{(++)}$ and $G_{(--)}(=G_{(++)}^{*})$
components, as well as the non-symmetric prescription for the $G_{(+-)}$
and $G_{(-+)}$ components.
The prescription is justified by the fact that the gluon radiations occur
definitely at positive $x^{-}$.

We note that there are the three temperature regimes:
(i) $T\ll P^{-}/x$,
(ii) $T\sim P^{-}/x$ and
(iii) $T\gg P^{-}/x$.
In the regime (i), thermal effects disappear up to LLA and the evolution
equation reduces to the vacuum BFKL equation. In the regime (iii),
the temperature is too high. We have a power enhancement rather than the
logarithmic one so that the evolution is described by an equation different
from the BFKL-type equation which resums the leading logarithmic terms.
This regime is beyond the scope of the present paper.
In the regime (ii), the temperature matches the light-cone energy of soft
gluons we are thinking of, giving rise to the logarithmic enhancement.
Thus, in this regime, it is meaningful to consider the thermal BFKL equation.
The resulting BFKL equation indicates that the thermal
effect shows up as the Bose enhancement of the soft gluon emissions due
to the surrounding thermal gluons. We have concluded that at finite
temperature, the growth of small $x$ gluons can be more rapid and the
saturation regime can be reached sooner than at zero temperature.

To complete, the full non-linear JIMWLK equation at finite temperature
will be derived in the future.

\begin{acknowledgments}
The authors are grateful to Larry McLerran and Kenji Morita for useful
discussions.
\end{acknowledgments}

\appendix

\section{Real-time thermal gluon propagator in light-front field theory}
\label{propagator}

In this appendix, we write down explicitly the real-time thermal gluon
propagator in light-front field theory with the light-cone gauge
($A_{a}^{+}=0$). Note the retarded prescription for $G^{i-}$ and $G^{-i}$
components.

\begin{align}
\rmi G^{i-}_{0(++)}(p) &= \frac{p^i}{p^{+}+\rmi\epsilon}
\left[
\rmi G_{0}(p)+2\pi n_{\rm B}\left(\left|p^{0}\right|\right)
\delta\left(p^{2}\right)
\right],
\nonumber\\
\rmi G^{i-}_{0(+-)}(p) &= \frac{p^i}{p^{+}+\rmi\epsilon}2\pi
\left[
\theta\left(-p^0\right)+ n_{\rm B}\left(\left|p^{0}\right|\right)
\right]
\delta\left(p^{2}\right),
\nonumber\\
\rmi G^{i-}_{0(-+)}(p) &= \frac{p^i}{p^{+}+\rmi\epsilon}2\pi
\left[
\theta\left(p^0\right)+ n_{\rm B}\left(\left|p^{0}\right|\right)
\right]
\delta\left(p^{2}\right),
\nonumber\\
\rmi G^{i-}_{0(--)}(p) &= \frac{p^i}{p^{+}-\rmi\epsilon}
\left[
-\rmi G_{0}^{*}(p)+2\pi n_{\rm B}\left(\left|p^{0}\right|\right)
\delta\left(p^{2}\right)
\right],
\end{align}
\begin{align}
\rmi G^{-i}_{0(++)}(p) &= \frac{p^i}{p^{+}-\rmi\epsilon}
\left[
\rmi G_{0}(p)+2\pi n_{\rm B}\left(\left|p^{0}\right|\right)
\delta\left(p^{2}\right)
\right],
\nonumber\\
\rmi G^{-i}_{0(+-)}(p) &= \frac{p^i}{p^{+}-\rmi\epsilon}2\pi
\left[
\theta\left(-p^0\right)+ n_{\rm B}\left(\left|p^{0}\right|\right)
\right]
\delta\left(p^{2}\right),
\nonumber\\
\rmi G^{-i}_{0(-+)}(p) &= \frac{p^i}{p^{+}-\rmi\epsilon}2\pi
\left[
\theta\left(p^0\right)+ n_{\rm B}\left(\left|p^{0}\right|\right)
\right]
\delta\left(p^{2}\right),
\nonumber\\
\rmi G^{-i}_{0(--)}(p) &= \frac{p^i}{p^{+}+\rmi\epsilon}
\left[
-\rmi G_{0}^{*}(p)+2\pi n_{\rm B}\left(\left|p^{0}\right|\right)
\delta\left(p^{2}\right)
\right],
\end{align}
\begin{align}
\rmi G^{ij}_{0(++)}(p) &= \delta^{ij}
\left[
\rmi G_{0}(p)+2\pi n_{\rm B}\left(\left|p^{0}\right|\right)
\delta\left(p^{2}\right)
\right],
\nonumber\\
\rmi G^{ij}_{0(+-)}(p) &= \delta^{ij} 2\pi
\left[
\theta\left(-p^0\right)+ n_{\rm B}\left(\left|p^{0}\right|\right)
\right]
\delta\left(p^{2}\right),
\nonumber\\
\rmi G^{ij}_{0(-+)}(p) &= \delta^{ij}2\pi
\left[
\theta\left(p^0\right)+ n_{\rm B}\left(\left|p^{0}\right|\right)
\right]
\delta\left(p^{2}\right),
\nonumber\\
\rmi G^{ij}_{0(--)}(p) &= \delta^{ij}
\left[
-\rmi G_{0}^{*}(p)+2\pi n_{\rm B}\left(\left|p^{0}\right|\right)
\delta\left(p^{2}\right)
\right],
\end{align}
\begin{align}
\rmi G^{--}_{0(++)}(p) &= {\rm PV}\frac{2p^-}{p^+}
\left[
\rmi G_{0}(p)+2\pi n_{\rm B}\left(\left|p^{0}\right|\right)
\delta\left(p^{2}\right)
\right],
\nonumber\\
\rmi G^{--}_{0(+-)}(p) &=  {\rm PV}\frac{2p^-}{p^+}2\pi
\left[
\theta\left(-p^0\right)+ n_{\rm B}\left(\left|p^{0}\right|\right)
\right]
\delta\left(p^{2}\right),
\nonumber\\
\rmi G^{--}_{0(-+)}(p) &= {\rm PV}\frac{2p^-}{p^+}2\pi
\left[
\theta\left(p^0\right)+ n_{\rm B}\left(\left|p^{0}\right|\right)
\right]
\delta\left(p^{2}\right),
\nonumber\\
\rmi G^{--}_{0(--)}(p) &= {\rm PV}\frac{2p^-}{p^+}
\left[
-\rmi G_{0}^{*}(p)+2\pi n_{\rm B}\left(\left|p^{0}\right|\right)
\delta\left(p^{2}\right)
\right],
\end{align}
where
\begin{align*}
G_{0}(p) &= \frac{1}{p^{2}+\rmi\epsilon},\\
n_{\rm B}(x) &= \frac{1}{{\rm e}^{\beta x}-1},\\
{\rm PV}\frac{1}{p^+} &\equiv\frac{1}{2}
\left(\frac{1}{p^{+}-\rmi\epsilon}+\frac{1}{p^{+}+\rmi\epsilon}\right).
\end{align*}


\end{document}